# MULTIMODAL SENSOR FUSION FOR REAL-TIME LOCATION-DEPENDENT DEFECT DETECTION IN LASER-DIRECTED ENERGY DEPOSITION

**Lequn Chen**
Advanced Remanufacturing and Technology Centre (ARTC), Singapore
Nanyang Technological University, Singapore

**Xiling Yao**
Singapore Institute of Manufacturing Technology (SIMTech), Singapore

**Wenhe Feng**
Advanced Remanufacturing and Technology Centre (ARTC), Singapore

**Youxiang Chew**
Advanced Remanufacturing and Technology Centre (ARTC), Singapore

**Seung Ki Moon**
Nanyang Technological University, Singapore

## ABSTRACT

*Real-time defect detection is crucial in laser-directed energy deposition (L-DED) additive manufacturing (AM). Traditional in-situ monitoring approach utilizes a single sensor (i.e., acoustic, visual, or thermal sensor) to capture the complex process dynamic behaviors, which is insufficient for defect detection with high accuracy and robustness. This paper proposes a novel multimodal sensor fusion method for real-time location-dependent defect detection in the robotic L-DED process. The multimodal fusion sources include a microphone sensor capturing the laser-material interaction sound and a visible spectrum CCD camera capturing the coaxial melt pool images. A hybrid convolutional neural network (CNN) is proposed to fuse acoustic and visual data. The key novelty in this study is that the traditional manual feature extraction procedures are no longer required, and the raw melt pool images and acoustic signals are fused directly by the hybrid CNN model, which achieved the highest defect prediction accuracy (98.5 %) without the thermal sensing modality. Moreover, unlike previous region-based quality prediction, the proposed hybrid CNN can detect the onset of defect occurrences. The defect prediction outcomes are synchronized and registered with in-situ acquired robot tool-center-point (TCP) data, which enables localized defect identification. The proposed multimodal sensor fusion method offers a robust solution for in-situ defect detection.*

## 1. INTRODUCTION

Laser-directed energy deposition (L-DED) is an additive manufacturing (AM) process that uses a laser beam as the heat source to melt metallic wire or powder and deposit material layer by layer onto a substrate. L-DED has been used in the aerospace, automotive, marine and offshore industries to produce complex large-scale metallic parts and components [1–4]. However, maintaining high quality consistency and process stability in such large-format metal AM process remains a serious challenge [5]. During the L-DED process, defects such as keyhole porosity and cracks could emerge stochastically due to heat accumulation, speed inconsistency and gas entrapment. Pores and cracks generated in the AM process can severely degrade the mechanical performance (e.g., strength, hardness, corrosion resistance) of printed parts. Therefore, early detection and correction of defects is critical for preventing build failures and ensuring as-built product quality.

Sensor-based in-situ monitoring is the key to identifying process anomalies and defects. The state-of-the-art in-situ monitoring techniques rely on a single sensor (i.e., either acoustic, visual, or thermal sensor) to capture the complex process dynamics. Vision-based melt pool monitoring is one of the most popular sensing approaches. For example, Scime and Beuth [6] used a high speed camera with a fixed field of view to capture melt pool morphologies in L-PBF. Keyhole porosity and balling instabilities can be identified using the melt pool images with supervised machine learning (ML) approaches. In addition, features extracted from the visible or infrared melt pool images such as melt pool width, size, and peak temperature can be used for closed-loop control of laser power to maintain process stability and enhance the microstructure homogeneity [7–9]. In-situ melt pool signatures can also be used for spatiotemporal modelling, process-structure-property (PSP) causal analytics, and design rule constructions in L-PBF [10–12]. Recently, we proposed an in-situ melt pool monitoring technique with a multi-feature extraction pipeline, using off-axis infrared thermal imaging [13].

Acoustic signal captured by a microphone sensor or Fiber Bragg Grating (FBG) sensor are yet another prominent in-situ sensing technique. For example, Surovi et al. [14] used features extracted from acoustic signals captured by a microphone to identify geometric anomalies in Wire Arc Additive Manufacturing (WAAM). Supervised ML models were applied to classify the sound generated from defective weld bead and non-defective weld bead. Similarly, Bevans et al. [15] applied a wavelet integrated graph method to extract features from acoustic signals to identify various types of flaws in wire-based DED, such as porosity, spatters, and line width variations. Acoustic based monitoring has also shown great interest in L-PBF, as presented in [16–18]. Recently, we introduced a signal denoising technique that can significantly enhance the acoustic-based defect prediction accuracy in L-DED process [19,20].

Surface monitoring can be achieved via laser line scanning and in-situ point cloud processing. In our previous work, a ML-assisted surface defect detection method was presented with in-situ 3D point cloud data feature extractions [21,22]. Surface anomalies and dimension deviations such as bulge and dent regions can be identified on-the-fly via supervised ML. Similar approach has been adopted in Fused Deposition Modelling (FDM) by Lyu et al. [23]. On top of the surface defect detection, an in-process adaptive dimension correction strategy was proposed in [24], where robotic AM toolpath was generated to automatically fill up the under-built area.

The above-mentioned single-modal sensing techniques show promise in predicting specific types of anomalies (e.g., pores, geometric defects, instability) for online monitoring of the metal AM processes. However, because most sensors could not adequately capture the complicated melt pool dynamics and laser-material interactions, single sensor-based approaches could not predict defects and part quality in a holistic and robust manner. Multimodal monitoring allows for a more comprehensive understanding of underlying complex physical phenomena. Different sensors could potentially alleviate each other's limitations, improving defect prediction accuracy and robustness [25]. Multimodal fusion has demonstrated promising results in several recent literature, outperforming single-modal defect predictions in the L-PBF process [26–28]. In our prior work, we extract handcrafted features from visual, acoustic, and thermal sensing modalities and developed a multisensor fusion-based digital twin (MFDT) for localized quality prediction in the robotic L-DED process [29,30]. In this paper, we propose a novel hybrid convolutional neural network (CNN) to fuse acoustic signals and coaxial melt pool images on top of the MFDT framework. The key novelty in this study is that the previous work's manual feature extraction procedures are no longer required, and the raw melt pool images and acoustic signals are fused directly by the hybrid CNN model, which achieved high defect prediction accuracy (98.5 %) without the thermal sensing modality. Moreover, unlike previous region-based quality prediction, the proposed hybrid CNN can detect the onset of defect occurrences. The defect prediction outcomes are synchronized and registered with in-situ acquired robot tool-center-point (TCP) position data, which enables location-specific defect identification.

## 2. MULTIMODAL DATASET

### 2.1 Experiments

Figure 1 depicts the setup of a coaxial visible spectrum CCD camera and microphone sensor. A near infrared (NIR) filter was attached to the lens of the coaxial camera, which was installed on the laser head. The microphone sensor was around 10 cm away from the process zone. Both sensors were connected to a PC running a customized software that extracted raw images and acoustic signals simultaneously. More information about the multisensor setup can be found in our previous publications [5,29,31].

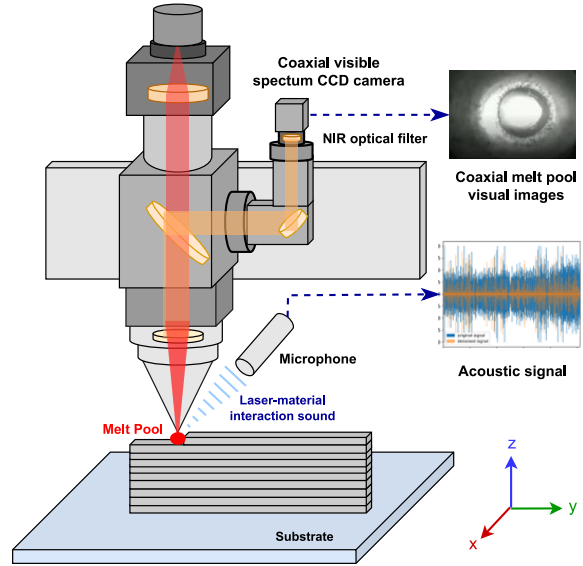

**FIGURE 1:** MULTIMODAL MONITORING EXPERIMENTAL SETUP IN THE L-DED PROCESS.

The multimodal in-situ monitoring dataset was created by depositing various single bead wall samples with maraging steel C300 powder. Table 1 summarizes the process parameters used during the experiments for multimodal data collection. In this study, we did not deliberately introduce defects by choosing suboptimal parameters. Instead, the process parameters were pre-optimized in order to attain near fully-dense quality for multi-layer multi-track industrial-level production. Due to the limited heat transfer capacities, localized heat accumulation became substantially worse when fabricating single-bead wall samples than when depositing block structures. As a result of the rapid heating and cooling cycles, defects such as cracks and keyhole pores emerged, as shown in the optical microscope (OM) image in Figure 2. After several layers of deposition, the transition from defect-free zone to defective zone occurs. We increased the dwell time between layers to postpone the emergence of defects since it provides cooling time to reduce localized heat build-up. As a result, defects appeared at higher layers for samples deposited with longer dwell times.

**TABLE 1:** L-DED PROCESS EXPERIMENTS FOR COLLECTING MULTIMODAL IN-SITU MONITORING DATASET.

| Parameters | Values |
|---|---|
| Laser power (kW) | 2.3-2.5 |
| Speed (mm/s) | 25-27.5 |
| Dwell time (s) | [0, 5, 10] |
| Laser beam diameter (mm) | 2 |
| Powder flow rate (g/min) | 12 |
| Hatch space (mm) | 1 |
| Layer thickness (mm) | 0.85 |
| Stand-off distance (mm) | 12 |

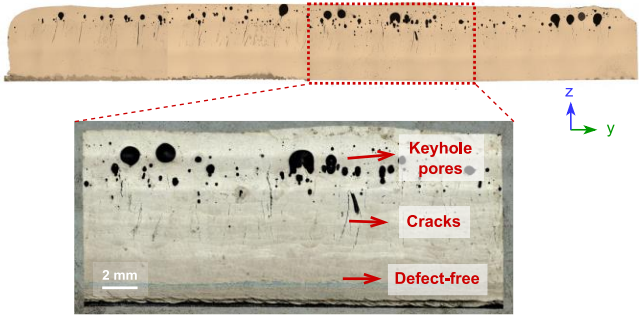

**FIGURE 2:** AN EXAMPLE OF THE OM IMAGE OF THE SINGLE BEAD WALL SAMPLE PRODUCED FOR MULTIMODAL DATA COLLECTION.

## 2.2 Dataset Descriptions

The multimodal dataset for training the deep learning model was constructed utilizing acoustic and vision sensor data collected from three single bead wall experiments. Each single bead wall consists of 50 layers. The acoustic signal and coaxial images were recorded simultaneously during the experiments. The melt pool images were acquired with a sampling rate of 10 Hz. Acoustic signals were denoised using the method described in our previous study [19]. The denoised audio signals are then segmented into 10 ms chunks in order to synchronize with the coaxial melt pool images for multimodal dataset development. Cracks and keyhole pores are both categorized as 'defective' in this study. Figures 3 and 4 illustrate examples of signals from different categories such as defective, laser-off, and defect-free printing status. Crack occurrences are characterized by sudden amplitude changes in the acoustic signal. This could be related to the energy released during crack propagation. As for the optical image, when cracks emerged, the melt pool became noticeably brighter and larger. Similarly, the amplitude of the acoustic signal in the keyhole pores region is larger than the defect-free region. The coaxial melt pool images with large variations exhibit greater and brighter melt pools, indicating severe process instabilities. The goal is to train supervised deep learning models on the collected multimodal dataset to differentiate between defect-free, defective, and laser-off events. Figure 5 depicts the multimodal dataset's distribution per category.

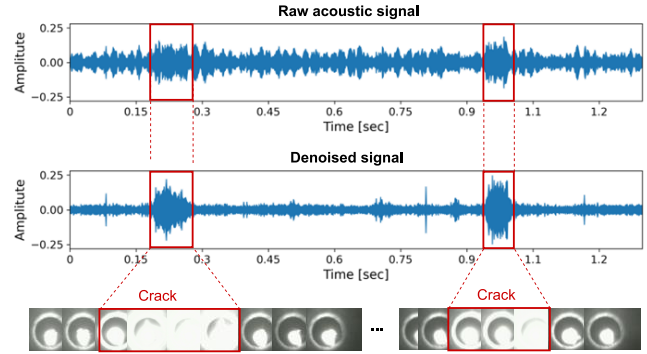

**FIGURE 3:** AN EXAMPLE OF CRACK EVENTS OBSERVED IN MELT POOL IMAGES AND ACOUSTIC SIGNALS. ABRUPT CHANGES CAN BE SEEN IN BOTH COAXIAL MELT POOL IMAGES AND DENOISED ACOUSTIC SIGNALS.

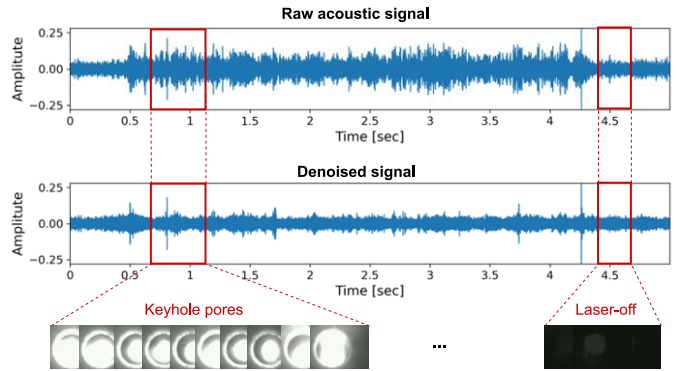

**FIGURE 4:** AN EXAMPLE OF KEYHOLE POROSITY REGIONS AND LASER-OFF REGIONS OBSERVED BY COAXIAL MELT POOL IMAGES AND ACOUSTIC SIGNALS. KEYHOLE POROSITY REGION HAS LARGER AND BRIGHTER MELT POOL AND HIGHER ACOUSTIC AMPLITUDE THAN DEFECT-FREE REGIONS.

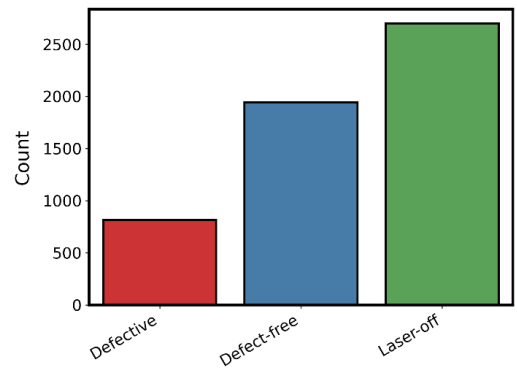

**FIGURE 5:** DISTRIBUTIONS OF L-DED MULTIMODAL DATASET PER CATEGORIES.

## 3. MULTIMODAL SENSOR FUSION

This section introduces the proposed multimodal sensor fusion method. First, a customized software platform is used to synchronize multimodal sensor data. The software offers an automated quality prediction pipeline that incorporates the trained CNN model for online quality prediction. Second, the proposed hybrid CNN model is compared to single-modal CNN architectures such as the conventional VGG19 model for image classification [32] and the MFCC-CNN model for L-DED acoustic signal classification presented in our earlier research [20].

## 3.1 Multimodal Sensor Data Synchronization, Registration, and Quality Prediction Pipeline

Figure 6 depicts the proposed software workflow for the multimodal sensor data synchronization, registration, and location-dependent quality prediction. The program was designed on the Ubuntu Linux operating system using the Robot Operating System (ROS). The software was built on top of our previous MFDT framework [29]. The key novelty in this study is that the manual feature engineering steps in our previous work are no longer needed, and the raw melt pool images and acoustic signals are fused directly by the hybrid CNN model. Moreover, the proposed hybrid CNN is able to identify onset of defect occurrences at 10 Hz, instead of previous region-based quality prediction.

Figure 6 depicts ROS nodes and ROS topics as blocks and circles, respectively. When the L-DED process starts, all of the ROS node applications run in parallel. Raw signals are extracted by the microphone sensor driver and the coaxial CCD camera driver and stored for offline data annotation and hybrid CNN model training. The trained hybrid CNN model is loaded into the software pipeline, which subscribes to the raw melt pool image and acoustic data at the same time. Based on the multimodal input, the hybrid CNN model infers the quality at each time stamp. The defect prediction frequencies are configured to be the same as the melt pool image acquisition frequency of 10 Hz. In contrast, the acoustic data is acquired at 44100 Hz. The raw acoustic signal is buffered and divided into 100 ms chunks with the same time stamp as the melt pool image data.

Meanwhile, the robot tool-center-point (TCP) position data is acquired via ethernet communication utilizing the KUKA Robot Sensor Interface (RSI). ROS can retrieve real-time robot TCP positions from the robot controller and publish them as a ROS topic. The quality prediction outcome (defective, defect-free, or laser-off) is registered with TCP position data (i.e., deposition toolpath x, y, z coordinates), resulting in location-specific defect predictions.

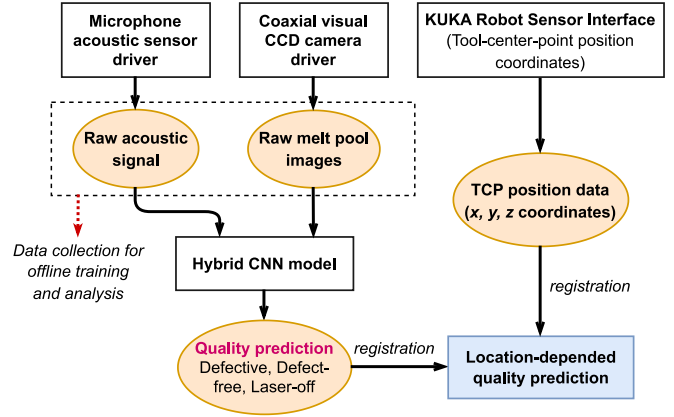

**FIGURE 6:** SOFTWARE WORKFLOW FOR REAL-TIME LOCATION-DEPENDENT QUALITY PREDICTION.

### 3.2 Hybrid CNN Model for Multimodal Sensor Fusion

This section introduces the proposed hybrid CNN model for multimodal sensor fusion. Furthermore, the multimodal in-situ monitoring dataset is used to benchmark single-modal CNN architectures such as the VGG 19 model for image classification and the MFCC-CNN for acoustic signal classification. Unlike single-modal CNN architectures, the proposed hybrid CNN model separates visual and audio models into independent streams and concatenates them to achieve feature-level fusions. The details are explained as follows.

VGG19 is a variant of VGG model [32] which consists of 19 layers (16 convolution layers, 3 Fully connected layer, 5 Maxpooling layers and 1 SoftMax layer). In this study, the implementation of VGG19 slightly differs than the original implementation, which does not include the fully connected layers. This is due to the fact that the original VGG19 was designed for ImageNet [33], where the images have RGB channels with complex features, whereas the melt pool images have gray scale images with less complex features. Fully connected layers are removed to prevent overfitting. The full architecture of modified VGG19 is shown in Figure 7, where the raw melt pool image of 640*480 pixels are transformed into 30*30 shape. Similarly, the MFCC-CNN model also adopts VGG-like architecture, where convolutional layers are connected with Maxpooling operations as shown in Figure 8. The input for the MFCC-CNN model is time-frequency representations of acoustic signal called Mel-frequency cepstrum coefficients (MFCCs). The MFCCs represents the acoustic signal information within 100 ms segments across the entire frequency bands. The details regarding MFCC-CNN model and hyperparameters can be found in our previous publication in [20].

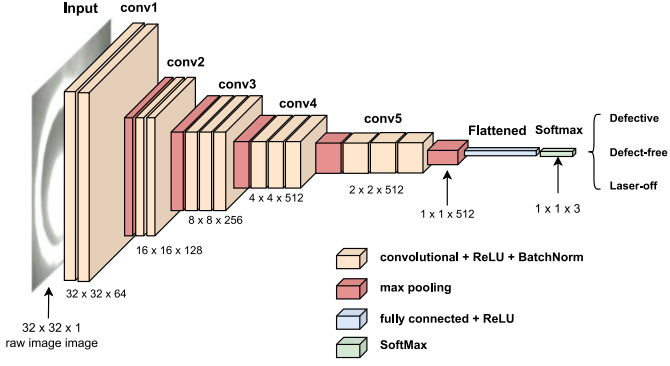

**FIGURE 7:** MODIFIED VGG19 ARCHITECTURE FOR COAXIAL MELT POOL IMAGE CLASSIFICATION.

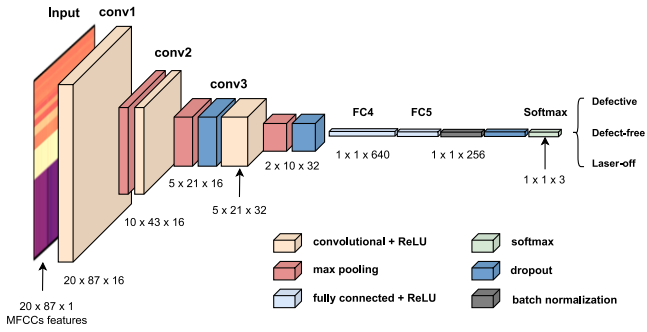

**FIGURE 8:** MFCC-CNN MODEL ARCHITECTURE FOR L-DED ACOUSTIC SIGNAL CLASSIFICATION [20].

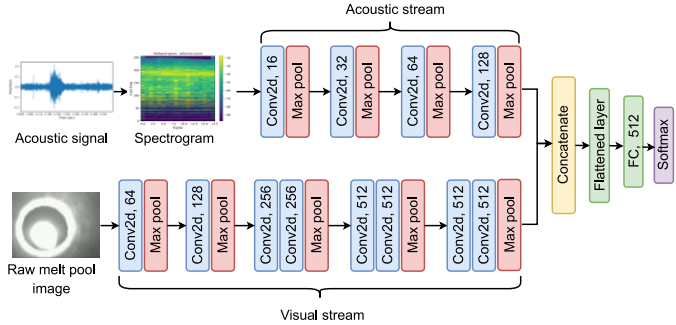

**FIGURE 9:** THE PROPOSED HYBRID CNN MODEL CONCATENATES THE IMAGE FEATURE EXTRACTION STREAM AND THE ACOUSTIC FEATURE EXTRACTION STREAM TO ACHIEVE FEATURE-LEVEL FUSION.

The proposed hybrid CNN for multimodal sensor fusion is shown in Figure 9. The hybrid CNN model consists of two streams: a feature extraction stream for acoustic signal and a feature extraction stream for melt pool image. The two streams use a VGG-like architecture. The visual stream has 8 convolutional layers and 5 Maxpooling layers, making it comparable to VGG11. The acoustic stream takes Mel-Spectrogram as input (i.e., logarithm scale of short-time Fourier transform) and feeds it into 4 convolutional layers and 4 Maxpooling layers. The acoustic and visual streams are merged and feed into a flattened layer. Finally, a fully connected layer with SoftMax is connected for multi-class classification. The PyTorch deep learning framework was used to implement all of the deep learning models. The next section will show the model performance evaluation on the collected dataset.

## 4. RESULTS AND DISCUSSION

The VGG19 model, MFCC-CNN model and proposed hybrid CNN model are trained and tested on the collected dataset. The dataset contains 5450 samples that are separated into a training set and testing set in an 8:2 ratio. Since the data distribution is imbalanced, the stratified random split was used to divide the training and testing datasets. The classification accuracies of VGG-19 trained on melt pool images, MFCC-CNN trained on acoustic data, and hybrid CNN trained on multimodal datasets are compared. The accuracy is defined by the following:

$$Accuracy = \frac{TP + TN}{TP + FP + TN + FN} \quad (1)$$

where TP, TN, FP, FN means true positive, true negative, false positive, and false negative, respectively. All three model evaluations reported in this paper were averaged across five runs to verify model viability and repeatability. Table 2 summarizes the mean accuracy and standard deviation of the accuracy on the testing set. With 98.53% mean accuracy and smaller standard deviations, the hybrid CNN model trained on multimodal dataset outperforms single-modal CNN models trained on individual dataset. This demonstrates that combining two sensor data sets can considerably improve defect prediction accuracy. Moreover, as illustrated in the confusion matrix in Figure 10, the proposed hybrid CNN model has better robustness for predicting defect-free class, while vision and acoustic-based models performed poorly.

It can be seen that the vision-based defect detection slightly outperforms the acoustic-based detection. However, as shown in the confusion matrix in Figure 10, the vision-based method demonstrates a lower performance in detecting the "defect-free" class (66.1%) compared to the acoustic-based monitoring method, which achieves an accuracy of 86.5%. Conversely, the vision sensor achieves a 100% accuracy in predicting the defective class, while the acoustic-based method exhibits a marginally lower performance (96.7%). This finding suggests that vision-based monitoring techniques are incapable of detecting subtle differences when the process transitions from defect-free to defective regions. Furthermore, the vision-based CNN tends to overestimate the occurrence of defects. This limitation stems from the inability of the visible spectrum camera to capture complicated melt pool dynamics. Acoustic sensors, on the other hand, may detect cracking or overheated sounds more easily. The multimodal fusion-based prediction achieves an accuracy of more than 90% for each of the three categories by fusing both sensing modalities. This implies that combining the two sensing modalities can help to overcome the limitations of each individual sensor, resulting in more accurate and reliable defect detection.

**TABLE 2:** TESTING ACCURACY OF SINGLE-MODAL PREDICTIONS AND MULTIMODAL PREDICTIONS

| Sensing modality | DL model | Testing accuracy (%) (Mean) | Testing accuracy (%) (std) |
|---|---|---|---|
| Acoustic | MFCC-CNN | 94.32 | 2.19 |
| Visual | VGG19 | 94.76 | 1.20 |
| Multimodal (acoustic + visual) | **Hybrid CNN (proposed method)** | **98.53** | **1.11** |

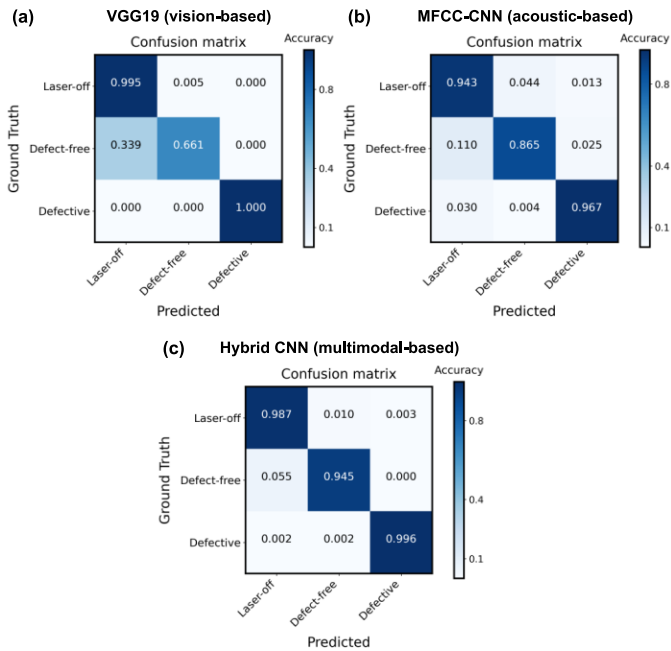

**FIGURE 10:** CONFUSION MATRIX. (A) VGG-19 PERFORMANCE ON VISION-ONLY DATASET, (B) MFCC-CNN PERFORMANCE ON ACOUSTIC-ONLY DATASET, (C) HYBRID CNN MODEL PERFORMANCE ON MULTIMODAL DATASET.

## 5. CONCLUSION

This paper proposed a novel hybrid CNN model to fuse raw melt pool image and acoustic data. The key novelty in this study is that the traditional manual feature extraction procedures are no longer needed, and raw melt pool images and acoustic signals are fused directly. With 98.5% accuracy, the hybrid CNN trained on the multimodal dataset surpasses the VGG19 model trained on melt pool images and MFCC-CNN trained on L-DED acoustic signals. In comparison to previous studies, the proposed technique obtained great accuracy without the use of an additional thermal sensing modality. Furthermore, instead of region-based defect prediction, the proposed hybrid CNN can detect the onset of defect occurrences. In addition, the defect prediction outcomes are synchronized and registered with in-situ acquired robot tool-center-point (TCP) data, allowing for localized defect identification. The proposed multimodal sensor fusion method offers a robust solution for in-situ defect detection.

## ACKNOWLEDGEMENTS

This research is funded by the Agency for Science, Technology and Research (A*STAR) of Singapore through the Career Development Fund (Grant No. C210812030). It is also supported by Singapore Centre for 3D Printing (SC3DP), the National Research Foundation, Prime Minister's Office, Singapore under its Medium-Sized Centre funding scheme.

## REFERENCES

[1] Tan, C., Weng, F., Sui, S., Chew, Y., and Bi, G., 2021, "Progress and Perspectives in Laser Additive Manufacturing of Key Aeroengine Materials," International Journal of Machine Tools and Manufacture, **170**, p. 103804.
[2] Su, J., Tan, C., Ng, F. L., Weng, F., Chen, L., Jiang, F., Teng, J., and Chew, Y., 2022, "Additive Manufacturing of Novel Heterostructured Martensite-Austenite Dual-Phase Steel through in-Situ Alloying," Materials Today Communications, **33**, p. 104724.
[3] Tan, C., Li, Q., Yao, X., Chen, L., Su, J., Ng, F. L., Liu, Y., Yang, T., Chew, Y., Liu, C. T., and DebRoy, T., 2023, "Machine Learning Customized Novel Material for Energy-Efficient 4D Printing," Advanced Science, p. 2206607.
[4] Lee, J., Chua, P. C., Chen, L., Ng, P. H. N., Kim, Y., Wu, Q., Jeon, S., Jung, J., Chang, S., and Moon, S. K., 2023, "Key Enabling Technologies for Smart Factory in Automotive Industry: Status and Applications," International Journal of Precision Engineering and Manufacturing-Smart Technology, **1**(1), pp. 93–105.
[5] Chen, L., Yao, X., Xu, P., Moon, S. K., Zhou, W., and Bi, G., 2023, "In-Process Sensing, Monitoring and Adaptive Control for Intelligent Laser-Aided Additive Manufacturing," *Transactions on Intelligent Welding Manufacturing*, Springer, Singapore, pp. 3–30.
[6] Scime, L., and Beuth, J., 2019, "Using Machine Learning to Identify In-Situ Melt Pool Signatures Indicative of Flaw Formation in a Laser Powder Bed Fusion Additive Manufacturing Process," Additive Manufacturing, **25**, pp. 151–165.

[7] Chen, L., Yao, X., Chew, Y., Weng, F., Moon, S. K., and Bi, G., 2020, "Data-Driven Adaptive Control for Laser-Based Additive Manufacturing with Automatic Controller Tuning," Applied Sciences, **10**(22), p. 7967.

[8] Smoqi, Z., Bevans, B. D., Gaikwad, A., Craig, J., Abul-Haj, A., Roeder, B., Macy, B., Shield, J. E., and Rao, P., 2022, "Closed-Loop Control of Meltpool Temperature in Directed Energy Deposition," Materials & Design, **215**, p. 110508.

[9] Freeman, F., Chechik, L., Thomas, B., and Todd, I., 2023, "Calibrated Closed-Loop Control to Reduce the Effect of Geometry on Mechanical Behaviour in Directed Energy Deposition," Journal of Materials Processing Technology, **311**, p. 117823.

[10] Ko, H., Lu, Y., Yang, Z., Ndiaye, N. Y., and Witherell, P., 2023, "A Framework Driven by Physics-Guided Machine Learning for Process-Structure-Property Causal Analytics in Additive Manufacturing," Journal of Manufacturing Systems, **67**, pp. 213–228.

[11] Ko, H., Kim, J., Lu, Y., Shin, D., Yang, Z., and Oh, Y., 2022, "Spatial-Temporal Modeling Using Deep Learning for Real-Time Monitoring of Additive Manufacturing," American Society of Mechanical Engineers Digital Collection.

[12] Ko, H., Witherell, P., Lu, Y., Kim, S., and Rosen, D. W., 2021, "Machine Learning and Knowledge Graph Based Design Rule Construction for Additive Manufacturing," Additive Manufacturing, **37**, p. 101620.

[13] Chen, L., Yao, X., Ng, N. P. H., and Moon, S. K., 2022, "In-Situ Melt Pool Monitoring of Laser Aided Additive Manufacturing Using Infrared Thermal Imaging," *2022 IEEE International Conference on Industrial Engineering and Engineering Management (IEEM)*, pp. 1478–1482.

[14] Surovi, N. A., Dharmawan, A. G., and Soh, G. S., 2021, "A Study on the Acoustic Signal Based Frameworks for the Real-Time Identification of Geometrically Defective Wire Arc Bead," American Society of Mechanical Engineers Digital Collection.

[15] Bevans, B., Ramalho, A., Smoqi, Z., Gaikwad, A., Santos, T. G., Rao, P., and Oliveira, J. P., 2023, "Monitoring and Flaw Detection during Wire-Based Directed Energy Deposition Using in-Situ Acoustic Sensing and Wavelet Graph Signal Analysis," Materials & Design, **225**, p. 111480.

[16] Wang, H., Li, B., and Xuan, F.-Z., 2022, "Acoustic Emission for in Situ Process Monitoring of Selective Laser Melting Additive Manufacturing Based on Machine Learning and Improved Variational Modal Decomposition," Int J Adv Manuf Technol, pp. 1–16.

[17] Tempelman, J. R., Wachtor, A. J., Flynn, E. B., Depond, P. J., Forien, J.-B., Guss, G. M., Calta, N. P., and Matthews, M. J., 2022, "Detection of Keyhole Pore Formations in Laser Powder-Bed Fusion Using Acoustic Process Monitoring Measurements," Additive Manufacturing, **55**, p. 102735.

[18] Drissi-Daoudi, R., Pandiyan, V., Logé, R., Shevchik, S., Masinelli, G., Ghasemi-Tabasi, H., Parrilli, A., and Wasmer, K., 2022, "Differentiation of Materials and Laser Powder Bed Fusion Processing Regimes from Airborne Acoustic Emission Combined with Machine Learning," Virtual and Physical Prototyping, **0**(0), pp. 1–24.

[19] Chen, L., Yao, X., and Moon, S. K., 2022, "In-Situ Acoustic Monitoring of Direct Energy Deposition Process with Deep Learning-Assisted Signal Denoising," Materials Today: Proceedings.

[20] Chen, L., Yao, X., Tan, C., He, W., Su, J., Weng, F., Chew, Y., Ng, N. P. H., and Moon, S. K., 2023, "In-Situ Crack and Keyhole Pore Detection in Laser Directed Energy Deposition through Acoustic Signal and Deep Learning," Additive Manufacturing, **69**, p. 103547.

[21] Chen, L., Yao, X., Xu, P., Moon, S. K., and Bi, G., 2020, "Surface Monitoring for Additive Manufacturing with In-Situ Point Cloud Processing," *2020 6th International Conference on Control, Automation and Robotics (ICCAR)*, pp. 196–201.

[22] Chen, L., Yao, X., Xu, P., Moon, S. K., and Bi, G., 2021, "Rapid Surface Defect Identification for Additive Manufacturing with In-Situ Point Cloud Processing and Machine Learning," Virtual and Physical Prototyping, **16**(1), pp. 50–67.

[23] Lyu, J., Akhavan Taheri Boroujeni, J., and Manoochehri, S., 2021, "In-Situ Laser-Based Process Monitoring and In-Plane Surface Anomaly Identification for Additive Manufacturing Using Point Cloud and Machine Learning," American Society of Mechanical Engineers Digital Collection.

[24] Xu, P., Yao, X., Chen, L., Zhao, C., Liu, K., Moon, S. K., and Bi, G., 2022, "In-Process Adaptive Dimension Correction Strategy for Laser Aided Additive Manufacturing Using Laser Line Scanning," Journal of Materials Processing Technology, **303**, p. 117544.

[25] Gutknecht, K., Cloots, M., Sommerhuber, R., and Wegener, K., 2021, "Mutual Comparison of Acoustic, Pyrometric and Thermographic Laser Powder Bed Fusion Monitoring," Materials & Design, **210**, p. 110036.

[26] Li, J., Zhou, Q., Cao, L., Wang, Y., and Hu, J., 2022, "A Convolutional Neural Network-Based Multi-Sensor Fusion Approach for in-Situ Quality Monitoring of Selective Laser Melting," Journal of Manufacturing Systems, **64**, pp. 429–442.

[27] Petrich, J., Snow, Z., Corbin, D., and Reutzel, E. W., 2021, "Multi-Modal Sensor Fusion with Machine Learning for Data-Driven Process Monitoring for Additive Manufacturing," Additive Manufacturing, **48**, p. 102364.

[28] Pandiyan, V., Masinelli, G., Claire, N., Le-Quang, T., Hamidi-Nasab, M., de Formanoir, C., Esmaeilzadeh, R., Goel, S., Marone, F., Logé, R., Van Petegem, S., and Wasmer, K., 2022, "Deep Learning-Based Monitoring of Laser Powder Bed Fusion Process on Variable Time-Scales Using Heterogeneous Sensing and Operando X-

Ray Radiography Guidance," Additive Manufacturing, **58**, p. 103007.

[29] Chen, L., Bi, G., Yao, X., Tan, C., Su, J., Ng, N. P. H., Chew, Y., Liu, K., and Moon, S. K., 2023, "Multisensor Fusion-Based Digital Twin for Localized Quality Prediction in Robotic Laser-Directed Energy Deposition," Robotics and Computer-Integrated Manufacturing (in-press).

[30] Chen, L., Yao, X., Liu, K., Tan, C., and Moon, S. K., 2023, "Multisensor Fusion-Based Digital Twin in Additive Manufacturing for in-Situ Quality Monitoring and Defect Correction."

[31] Xu, P., Yao, X., Chen, L., Liu, K., and Bi, G., 2020, "Heuristic Kinematics of a Redundant Robot-Positioner System for Additive Manufacturing," *2020 6th International Conference on Control, Automation and Robotics (ICCAR)*, pp. 119–123.

[32] Simonyan, K., and Zisserman, A., 2015, "Very Deep Convolutional Networks for Large-Scale Image Recognition," arXiv:1409.1556 [cs].

[33] KrizhevskyAlex, SutskeverIlya, and E, H., 2017, "ImageNet Classification with Deep Convolutional Neural Networks," Communications of the ACM.